\documentclass[preprint,pra,showpacs,nofootinbib]{revtex4}
\usepackage[mathscr]{euscript}
\usepackage[dvips]{pstcol}
\usepackage{graphicx}
\usepackage{float,epsfig}
\usepackage{dcolumn}
\usepackage{bm}
\usepackage{graphicx}
\usepackage{dcolumn}
\usepackage{amsmath,amssymb,amsthm}
\usepackage{subfigure}
\usepackage[colorlinks=true,linkcolor=blue]{hyperref}
\newcommand{\bea}{\begin{eqnarray}}
\newcommand{\eea}{\end{eqnarray}}
\newcommand{\beq}{\begin{equation}}
\newcommand{\eeq}{\end{equation}}

\def\/{\over}

\begin{document}

\title{Electromagnetic shielding in quantum metrology}
\author{  Yao Jin$^{1}$  and Hongwei Yu$^{1,2,}$\footnote{Corresponding author} }
\affiliation{$^1$ Institute of Physics and Key Laboratory of Low
Dimensional Quantum Structures and Quantum
Control of Ministry of Education,\\
Hunan Normal University, Changsha, Hunan 410081, China \\
$^2$ Center for Nonlinear Science and Department of Physics, Ningbo
University, Ningbo, Zhejiang 315211, China}


\begin{abstract}
 The dynamics of the quantum Fisher information of the parameters of the initial atomic state and atomic transition frequency is
 studied, in the framework of open quantum systems,  for a static polarizable two-level atom coupled in the multipolar scheme to a bath of fluctuating vacuum electromagnetic fields without and with the presence of a reflecting boundary. Our results show that in the case without a boundary, the
 electromagnetic vacuum fluctuations always cause the quantum Fisher information of the initial parameters and thus the precision limit of parameter estimation  to decrease. Remarkably, however, with the presence of a boundary, the quantum Fisher information becomes position  and atomic polarization dependent, and  as a result, it may be enhanced as compared to that in the case without a boundary and may even be shielded from the influence of the vacuum fluctuations in certain circumstances as if it were a closed system.

\end{abstract}

\maketitle

\section{Introduction}

In estimation theory, the Cram\'{e}r-Rao bound~\cite{Helstrom,Holevo} was  proposed to describe how well one can estimate a parameter from the probability distribution and the Fisher information is used to describe the precision limit.   Since quantum mechanics is strongly related to probability theory in the sense that when we make quantum measurements on quantum mechanical systems, the observed outcomes follow a probability distribution, the Fisher information is readily extended to quantum regime and the inverse of the so-called quantum Fisher information (QFI) gives the lower bound of the error of the parameter estimation in quantum metrology~\cite{Helstrom,Holevo,Hubner,Braunstein}. With different models of the probe systems and different parameters to be estimated, many applications of quantum metrology have been done and some of those are of practical significance in quantum technology such as quantum frequency standards~\cite{Bollinger}, optimal quantum clock~\cite{Buzek}, measurement of gravity accelerations~\cite{Peters}, clock synchronization~\cite{Jozsa}, only to name a few.  The central task in quantum metrology is to improve the precision of parameter estimation. Since a larger quantum Fisher information means better precision,  increasing the QFI becomes a key issue in quantum
metrology. A straightforward  way,  in addition to increasing the QFI, to enhance the  precision when the
probe systems are closed  is by parallel measurements.  Later it was shown that the use of correlated  systems such as entangled states can also improve the
precision of parameter estimation\cite{Bollinger,Yurke,Dowling,Kok,Giovannetti04,Giovannetti06,Boixo,Roy,Boixo08,Hofmann,Esteve,Hyllus10,Pezze,Hyllus}.  On the other hand , however, interaction between a system and an environment is unavoidable in reality, and the quantum decoherence induced by such interactions may decrease the
QFI and destroy the quantum entanglement in the probe system exploited to improve the precision. In this regard,
 It has been shown that  the  interaction between a system and an environment usually makes the measurements noisy, which in turn degrades the estimation precision~\cite{Rosenkranz,Huelga,Ulam-Orgikh,Sasaki,Shaji,Monras,Demkowicz12,Demkowicz,Lee,Dorner,Watanabe,Knysh,koldy,Kacprowicz,Genoni,Chin,Escher,Ma,Zhong,Chaves}.

One environment which no system can be isolated from is the vacuum that fluctuates all the time in quantum sense.  In the present paper, we are interested in how the vacuum fluctuations affect the QFI  regarding the estimation of  parameters of the initial states of a probe system which is modeled by a neutral polarizable two-level atom interacting with electromagnetic vacuum fluctuations. We first examine how the decoherence caused by the vacuum fluctuations decreases the QFI as time evolves in an unbound space, then we ask what happens if the
vacuum fluctuations are changed somehow, for example, by the presence of a reflecting boundary, i.e, can we protect the QFI from decreasing?  We demonstrate that in an unbounded space,  the QFI decreases exponentially  with time. This behavior is similar to that observed in the amplitude-damping channel of decoherence derived in~\cite{Zhong}. However, with the presence of a boundary, the situation changes dramatically in certain circumstances such that the QFI may even be protected from decreasing as if it were isolated from the environment. In other words, QFI may be shielded from quantum decoherence due to electromagnetic
vacuum fluctuations in some special cases. At this point, let us note that the modification of vacuum fluctuations has been demonstrated to
yield many interesting quantum phenomena such as the Casimir effect~\cite{Casimir,CasimirP}, the light-cone fluctuations when gravity is quantized~\cite{Yu99}, the Brownian motion of test particles in an electromagnetic vacuum~\cite{Yu}, the position-dependent spontaneous decay rates and geometric phase  in an electromagnetic vacuum~\cite{Yu06,Yu12}.

\section{quantum fisher information and dynamical evolution of a two-level atom coupled with vacuum fluctuations}

For a given quantum state $\rho(X)$ parametrized by an unknown parameter $X$, 
the unknown parameter can be inferred from a set of measurements, usually modelled mathematically by a set of positive-operator valued measures, on the state.  After the optimization of the measurements and the estimator, a precision limit of the unknown parameter estimation is obtained~\cite{Braunstein}
\beq
Var(X)\geq\frac{1}{N F_X}\;,\label{Un}
\eeq
where $N$ represents the repeated times and $F_X$ denotes the quantum Fisher information of parameter $X$ given by
\beq
F_X=\textrm{Tr}\,(\rho(X)L^2)\;.
\eeq
Here $L$ is the so-called  symmetric logarithmic derivative Hermitian operator, which satisfies the equation $\partial_X \rho(X)=\frac{1}{2}\{\rho(X),L\}$ with $\{\}$ standing for the anti-commutator. In the orthonormal basis $\rho(X)=\sum_{i=1}^N p_i|\psi_i\rangle\langle\psi_i|$, $L$ can be solved for and the quantum Fisher information  can be written as~\cite{Braunstein}
\beq\label{FI}
F_X=2\sum_{m,n}^N\frac{|\langle\psi_m|\partial_X\rho|\psi_n\rangle|^2}{p_m+p_n}\;.
\eeq
For a two-level system, the state of the system can be expressed in the Bloch sphere representation as
\beq
\rho=\frac{1}{2}(I+\bm{\omega}\cdot\bm{\sigma})\;,\label{Bloch0}
\eeq
where $\bm{\omega}=(\omega_1,\omega_2,\omega_3)$ is the Bloch vector and $\bm{\sigma}=(\sigma_1,\sigma_2,\sigma_3)$ denotes the Pauli matrices. As a result, $F_X$ can be expressed in a simple form~\cite{Zhong}
\begin{equation}\label{FQ}
   F_X=\left\{
    \begin{array}{l}
    \overset{.}|\partial_X\bm{\omega}|^2+\frac{(\bm{\omega}\cdot\partial_X\bm{\omega})^2}{1-|\bm{\omega}|^2}\;,\;\;\,|\bm{\omega}|<1\;,  \\
    |\partial_X\bm{\omega}|^2\;,\;\;\;\;\;\;\;\;\;\;\;\;\;\;\;\;\;\;\; |\bm{\omega}|=1\;. \\
    \end{array}
    \right.
   \end{equation}
Now let us first calculate the QFI for the initial state' s parameter estimation for an arbitrary state of the two-level atom
\beq
|\psi\rangle=\cos\frac{\theta}{2}|+\rangle+e^{i\phi}\sin\frac{\theta}{2}|-\rangle\;,\label{initial}
\eeq
where $\theta$ and $\phi$ correspond to the weight parameter and phase parameter, $|+\rangle$, $|-\rangle$ denote the excited state and ground state of the atom respectively, and the Bloch vector of the state can be represented as $\bm{\omega}=(\sin\theta\cos\phi,\sin\theta\sin\phi,\cos\theta)$. So the quantum Fisher information of $\theta$ and $\phi$ can be easily calculated as $F_\theta=1$ and $F_\phi=\sin^2\theta$. Taking the atom as a closed system, whose evolution is governed by the Hamiltonian $H_s={1\over
2}\,\hbar\omega_0\sigma_3$, with $\omega_0$ denoting  the transition frequency,  one  can easily show that the Bloch vector of the state with time $\tau$ becomes $\bm{\omega}=(\sin\theta\cos(\phi+\omega_0\tau),\sin\theta\sin(\phi+\omega_0\tau),\cos\theta)$, and the quantum Fisher information remains $F_\theta=1$ and $F_\phi=\sin^2\theta$, which is time independent. Thus the unitary evolution does not change the quantum Fisher information of initial parameters.

 However, if the interaction with an environment is considered, the story may be different  and the influence  of the  environment will in general be encoded in the atomic state with time. This is just what we are going to discuss next, and we consider how the QFI
 changes when the interaction with electromagnetic vacuum fluctuations is taken into account. For this purpose, let us study a  static polarizable two-level  atom
interacting with fluctuating electromagnetic fields in vacuum, and in this case,
the total Hamiltonian of the coupled system can be written as
$
H=H_s+H_f+H'
$,
where $H_s$, which is given before, is the Hamiltonian of
the atom. $H_f$ denotes the
Hamiltonian of the free electromagnetic field and its explicit expression
is not required here. The Hamiltonian that describes the interaction
between the atom and the  electromagnetic field can be written in the multipolar
coupling scheme  as
\beq\label{HI}
 H'(\tau)=-e\textbf{r} \cdot
\textbf{E}(x(\tau))\;,
\eeq
where {\it e} is the electron
electric charge, $e\,\bf r$  is the atomic electric dipole moment, and
${\bf E}(x)$ denotes the electric field strength. Let us note that, since both $\textbf{r}(\tau)$ and
$\textbf{E}(x)$ are not world vectors,  the interaction
Hamiltonian $H' $ is ambiguous when we deal with  atoms in motion. However, a manifestly coordinate invariant
generalization of $H' $ can be given \cite{Tak}:
\begin{equation}
H'(\tau)=-e\,r^{\mu}(\tau)\,F_{\mu\nu}(x(\tau))\,u^{\nu}(\tau)\;,
\;\label{HI4}
\end{equation}
where $F_{\mu\nu}$ is the field strength, $r^{\mu}(\tau)$ is a
four-vector and its temporal component in the frame of the
atom vanishes and its spatial components
in the same frame are given by $\textbf{r}(\tau)$, and $u^{\nu}$
is the four velocity of the atom. Since we have $u^{\nu}(\tau)=(1,0,0,0)$
in the frame of the atom, this extended interaction Hamiltonian
reduces to that given by Eq.~(\ref{HI}) in the reference frame of
the atom.  So we choose to work in this reference
frame. Notice that  we assume that the atom is static, as a result,  the laboratory frame is equivalent to the frame of the atom.

We let $\rho_{tot}=\rho(0) \otimes |0\rangle\langle0|$  be the initial total density matrix of the system. Here $\rho(0)$ is the initial reduced density matrix of the atom, which corresponds to the atomic state in Eqs.~(\ref{initial}), and $|0\rangle$ is the vacuum state of the field. The evolution of the total density matrix $\rho_{tot}$ in the proper time $\tau$ reads as
\begin{equation}\label{evo}
\frac{\partial\rho_{tot}(\tau)}{\partial\tau}=-{i\/\hbar}[H,\rho_{tot}(\tau)]\;.
\end{equation}
We assume that the interaction between the atom and  field is
weak. So, the evolution of the reduced
density matrix $\rho(\tau)$ can be written in the
Kossakowski-Lindblad form~\cite{Lindblad, pr5}
\begin{equation}\label{master}
{\partial\rho(\tau)\over \partial \tau}= -{i\/\hbar}\big[H_{\rm eff},\,
\rho(\tau)\big]
 + {\cal L}[\rho(\tau)]\ ,
\end{equation}
where
\begin{equation}
{\cal L}[\rho]={1\over2} \sum_{i,j=1}^3
a_{ij}\big[2\,\sigma_j\rho\,\sigma_i-\sigma_i\sigma_j\, \rho
-\rho\,\sigma_i\sigma_j\big]\ .
\end{equation}
The coefficients of the Kossakowski matrix $a_{ij}$ can be expressed as
\begin{equation}
a_{ij}=A\delta_{ij}-iB
\epsilon_{ijk}\delta_{k3}-A\delta_{i3}\delta_{j3}\;,
\end{equation}
with
\begin{equation}\label{abc}
A=\frac{1}{4}[{\cal {G}}(\omega_0)+{\cal{G}}(-\omega_0)]\;,\;~~
B=\frac{1}{4}[{\cal {G}}(\omega_0)-{\cal{G}}(-\omega_0)]\;.
\end{equation}
We define a  two-point correlation function, $G^{+}(x-x')$, which is related to the two-point functions of the electromagnetic fields, $\langle0|E_i(x)E_j(x')|0 \rangle$, as
\begin{equation}
G^{+}(x-x')={e^2\/\hbar^2} \sum_{i,j=1}^3\langle +|r_i|-\rangle\langle -|r_j|+\rangle\,\langle0|E_i(x)E_j(x')|0 \rangle\;,
\end{equation}
and its Fourier and Hilbert transforms,  ${\cal G}(\lambda)$ and ${\cal K}(\lambda)$,  then follows
\begin{equation}
{\cal G}(\lambda)=\int_{-\infty}^{\infty} d\Delta\tau \,
e^{i{\lambda}\Delta\tau}\, G^{+}\big(\Delta\tau\big)\; ,
\quad\quad
{\cal K}(\lambda)=\frac{P}{\pi
i}\int_{-\infty}^{\infty} d\omega\ \frac{ {\cal G}(\omega)
}{\omega-\lambda} \;.
\end{equation}
By absorbing the Lamb shift term, the effective Hamiltonian $H_{\rm eff}$ can be written as
\begin{equation}\label{heff}
H_{\rm eff}=\frac{1}{2}\hbar\Omega\sigma_3={\hbar\over 2}\{\omega_0+{i\/2}[{\cal
K}(-\omega_0)-{\cal K}(\omega_0)]\}\,\sigma_3\;,
\end{equation}
where $\Omega$ is the effective level spacing of the atom.
By applying Eq.~(\ref{Bloch0}) to Eq.~(\ref{master}), the Bloch vector with proper time $\tau$ can be solved as:
\bea\label{Bloch}
&&\omega_1(\tau)=\sin\theta\cos(\Omega\tau+\phi)\,e^{-2A\tau}\;,\nonumber\\
&&\omega_2(\tau)=\sin\theta\sin(\Omega\tau+\phi)\,e^{-2A\tau}\;,\\
&&\omega_3(\tau)=\cos\theta\, e^{-4A\tau}-\frac{B}{A}(1-e^{-4A\tau})\nonumber\;.
\eea


\section{Influence of vacuum fluctuations on initial parameter estimation}
Let us now examine how the vacuum fluctuations affect the quantum Fisher information and thus the precision of the initial parameter estimation. Using the  following electric two-point function~\cite{Greiner}
\begin{eqnarray}\label{2p2}
\langle0|E_i(x(\tau))E_j(x(\tau'))|0\rangle_0&=&{\hbar c\/4\pi^2\varepsilon_0}(\partial
_0\partial_0^\prime\delta_{ij}-\partial_i\partial_j^\prime)\nonumber\\&&\times
{1\/(x-x')^2+(y-y')^2+(z-z')^2-(ct-ct'-i\varepsilon)^2}\;,\label{ee1}
\end{eqnarray}
where
$\varepsilon\rightarrow+0$, $\partial^\prime$ denotes the
differentiation with respect to $x^\prime$ and the subscript $0$ indicates the vacuum two-point functions in the unbounded space. Applying the trajectory of the  atom
\begin{eqnarray}\label{traj}
t(\tau)=\tau\;, \ \ \ x(\tau)=x_0\;, \ \ \
y(\tau)=y_0\;,\ \ \ z(\tau)=z_0\;,
\end{eqnarray}
we find that the electric-field two-point functions can be written as
\beq
\langle0|E_i(x(\tau))E_j(x(\tau'))|0\rangle_0
={\hbar c\/\pi^2\varepsilon_0}{\delta_{ij}\/(c\,\Delta\tau-i\varepsilon)^4}\;,
\eeq
So the Fourier transform of the correlation functions can be calculated as
\beq\label{fourier0}
{\cal G}^{(0)}(\lambda)=\sum_i{e^2|\langle -|r_i|+\rangle|^2\lambda^3\/3\pi\varepsilon_0\hbar c^3}\theta(\lambda)\;,
\eeq
with $\theta(\lambda)$ being the standard step function. Let us note that here we let $\varepsilon=0$ after the calculation.
 The coefficients of the Kossakowski matrix $a_{ij}$ and the effective level spacing of the atom are now given by
\beq
A^{(0)}=B^{(0)}={\gamma_0\/4}\;,
\eeq
\beq\label{lm}
\Omega^{(0)}=\omega_0+{\gamma_0\/2\pi\omega_0^3}\,P\int_0^\infty
d\omega\,\omega^3\bigg({1\/\omega+\omega_0}-{1\/\omega-\omega_0}\bigg)\,\;. \eeq
Here $\gamma_0=e^2|\langle -|{\bf
r}|+\rangle|^2\,\omega_0^3/3\pi\varepsilon_0\hbar c^3$ denotes the spontaneous emission rate in vacuum without boundaries,
As a result, the quantum Fisher information of the initial weight and phase parameter become
\beq
F_\phi^{(0)}=\sin^2\theta\; e^{- \gamma_0\tau}
\eeq
and
\beq
F_\theta^{(0)}=e^{- \gamma_0\tau}\;.
\eeq
This shows that the QFI of both weight and phase parameters decreases exponentially with time due to the decoherence caused by the interaction between the atom and the fluctuating vacuum and the decay rate equals the spontaneous emission rate of the atom in vacuum.
 Therefore, when the measurement time is larger than the relaxation time $1/\gamma_0$, the precision of the estimation is greatly damaged. This kind behavior of the QFI is the same as that in the amplitude-damping channel described in~\cite{Zhong} as expected.


Since vacuum fluctuation will be modified if we set a boundary in the vacuum, we may wonder how the presence of  a boundary which confines the vacuum fluctuations of the field affects the quantum Fisher information of the initial parameters with time.
Now we consider the case  with the presence of a reflecting boundary at $z=0$~\footnote{Let us note here that external forces may be needed to balance the Casimir-Polder force to keep a polarizable atom at a fixed position near a reflecting boundary. }.  The electric two-point functions in this case can be expressed as a sum of the Minkowski vacuum term and a
correction term due to the boundary:
\begin{eqnarray}\label{2p1}
\langle E_i(x(\tau))E_j(x(\tau'))\rangle=\langle
E_i(x(\tau))E_j(x(\tau'))\rangle_0 +\langle
E_i(x(\tau))E_j(x(\tau'))\rangle_b\;,
\end{eqnarray}
where
the $\langle
E_i(x(\tau))E_j(x(\tau'))\rangle_0$ is the two-point function in the unbounded vacuum which has already been calculated  above and
\begin{eqnarray}\label{2p3}
\langle0|
E_i(x(\tau))E_j(x(\tau'))|0\rangle_b&=&-{\hbar c\/4\pi^2\varepsilon_0}[\,(\delta_{ij}-2n_in_j)\,\partial
_0\partial_0^\prime-\partial_i\partial_j^\prime\,]\nonumber\\&&\times
{1\/(x-x^\prime)^2+(y-y^\prime)^2+(z+z^\prime)^2-(ct-ct^\prime-i\varepsilon)^2}\;\label{ee2}
\end{eqnarray}
gives the correction due to the boundary. Here $\mathbf{n}=(0,0,1)$ is the unit vector normal to the boundary.
By applying the trajectory in Eq.~(\ref{traj}), the boundary term of the electric-field two-point functions can be written as
\bea
&&\langle0|E_x(x(\tau))E_x(x(\tau'))|0\rangle_b=\langle0|E_y(x(\tau))E_y(x(\tau'))|0\rangle_b
=-{\hbar c\/\pi^2\varepsilon_0}{c^2\Delta\tau^2+4z_0^2\/[\;(c\,\Delta\tau-i\varepsilon)^2-4z_0^2]^3}\;,\nonumber\\
&&\langle0|E_z(x(\tau))E_z(x(\tau'))|0\rangle_b
={\hbar c\/\pi^2\varepsilon_0}{1\/[\;(c\,\Delta\tau-i\varepsilon)^2-4z_0^2]^2}\;.
\eea
The Fourier transform of the correlation functions can be calculated as~\cite{Yu12}:
\beq\label{fourier}
{\cal G}(\lambda)=\sum_i{e^2|\langle -|r_i|+\rangle|^2\lambda^3\/3\pi\varepsilon_0\hbar c^3}(1-f_i(\lambda,z_0))\theta(\lambda)\;,
\eeq
with
\beq\label{fourier1}
f_x(\lambda,z_0)=f_y(\lambda,z_0)={3c^3\/16\lambda^3z_0^3}\bigg[{2\lambda z_0\/c}\cos{2\lambda z_0\/c}+\bigg({4\lambda^2z_0^2\/c^2}-1\bigg)\sin{2\lambda z_0\/c}\bigg]\;,
\eeq
\beq\label{fourier2}
f_z(\lambda,z_0)={3c^3\/8\lambda^3z_0^3}\bigg[{2\lambda z_0\/c}\cos{2\lambda z_0\/c}-\sin{2\lambda z_0\/c}\bigg]\;.
\eeq
Thus the coefficients of the Kossakowski matrix $a_{ij}$ and the effective level spacing of the atom now become
\beq
A=B={\gamma_0\/4}(1-\sum_i\alpha_if_i(\omega_0,z_0))\;,
\eeq
\beq\label{lm}
\Omega=\omega_0+{\gamma_0\/2\pi\omega_0^3}\,P\int_0^\infty
d\omega\,\omega^3\big(1-\sum_i\;\alpha_if_i(\omega_0,z_0)\big)\bigg({1\/\omega+\omega_0}-{1\/\omega-\omega_0}\bigg)\,\;, \eeq
where $\alpha_i=|\langle
-|r_i|+\rangle|^2/|\langle -|{\bf r}|+\rangle|^2\,.$ Physically, $\alpha_i$ represents the relative polarizability and they satisfy $\sum_i\alpha_i=1$.
We let $f(\omega_0,z_0)\equiv\sum_i\alpha_if_i(\omega_0,z_0)$. As a result, the quantum Fisher information of the initial weight and phase parameters can be expressed as follows
\beq
F_\phi=\sin^2\theta\; e^{- \gamma_0(1-f(\omega_0,z_0))\tau}
\eeq
and
\beq
F_\theta=e^{- \gamma_0(1-f(\omega_0,z_0))\tau}\;.
\eeq
So, the decay rate of the QFI is now modified by the factor, $1-f(\omega_0,z_0)$, as compared to the unbounded case. Let us now first examine the asymptotic behaviors of the QFI, that is
when the atom is placed very far from the boundary or very close to it.
When $z_0\rightarrow\infty$, $f_i(\omega_0,z_0)\rightarrow0$, the unbounded case is recovered as expected. When $z_0\rightarrow0$, $f_x(\omega_0,z_0)=f_y(\omega_0,z_0)=-f_z(\omega_0,z_0)=1$, so atoms with different polarization will behave differently.  When the polarization is along the $z$-axis, i.e., $\alpha_x=\alpha_y=0$, the decay rate becomes double of that in the unbounded Minkowski vacuum case, which makes the QFI decay even faster than that without the boundary. However, when the polarization  is in the $xy$ plane, i.e., $\alpha_z=0$, the decay rate becomes zero, which means that the QFI is totally protected from electromagnetic vacuum fluctuations for transversely polarizable atoms extremely close to the boundary as if it were isolated.  For an isotropic polarization, $\alpha_x=\alpha_y=\alpha_z=\frac{1}{3}$, the decay rate is $\frac{2}{3}\gamma_0$, so  the QFI decreases slower and the precision is enhanced by the presence of the boundary as compared to the
unbounded case.

\begin{figure}[htbp]
\centering
\includegraphics[height=2.1in,width=2.1in]{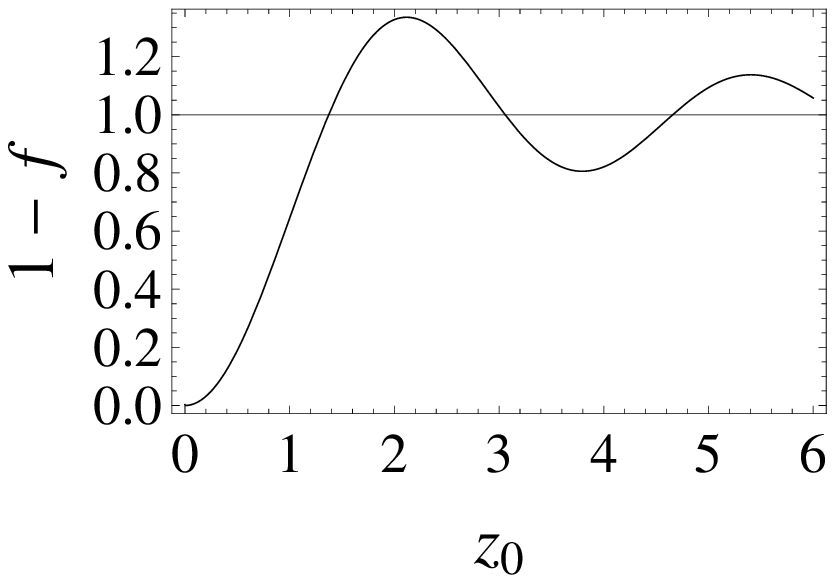}
\includegraphics[height=2.1in,width=2.1in]{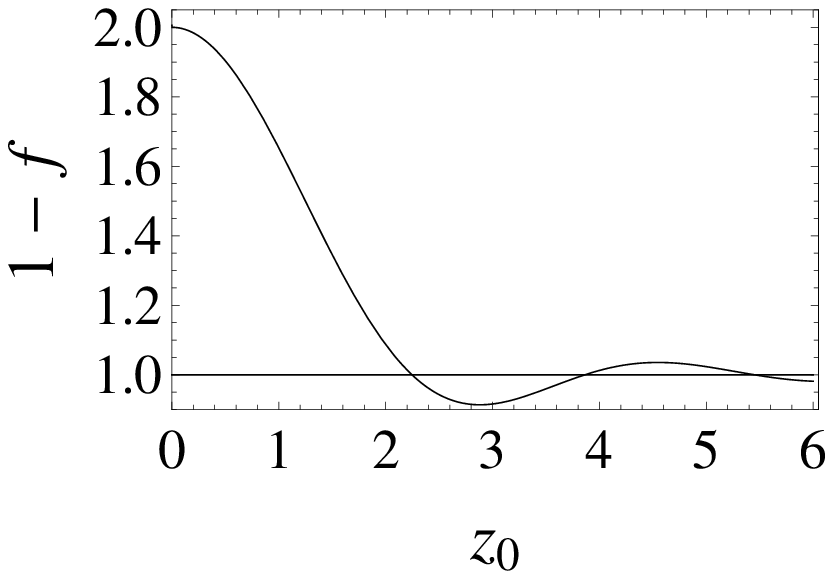}
\includegraphics[height=2.1in,width=2.1in]{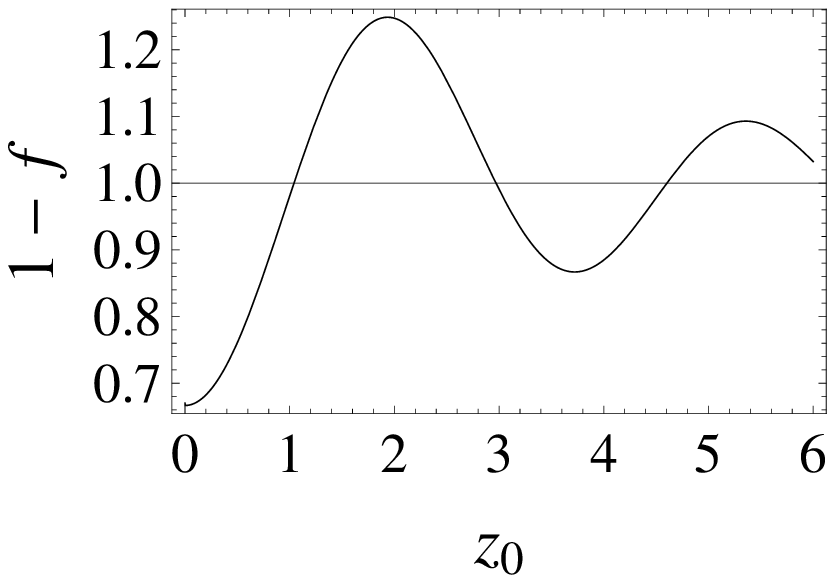}
\caption{ $1-f$ as  a function of $z_0$ for $\bm{\alpha}=(1,0,0),(0,0,1),(1/3,1/3,1/3)$ respectively. Here $z_0$ is in the unit of $c/\omega_0$. 
}\label{f}
\end{figure}

For a generic position and polarization, the QFI may  be decreased, or enhanced  as compared to the unbounded case.
This can be seen from the fact that $f_x$ decays in an oscillatory manner  from 1 to 0 with increasing atomic distance, while $f_z$ does from $-1$ to $0$.
In general, there exist some special positions
where $\sum_i\alpha_if_i(\omega_0,z_0)=0$. The quantum Fisher information of the initial parameters of the atom at these positions takes the same form as that in the unbounded case and the boundary effects disappear.  In  regions between these special positions,  we have either $\sum_i\alpha_if_i(\omega_0,z_0)>0$ or $<0$. As a result, the quantum Fisher information is either enhanced or decreased as compared to that in the unbounded space. To show the properties we described above graphically, we plot,
in Fig.~(\ref{f}),  $1-f(\omega_0,z_0)$ as a function of $z_0$ for
$\bm{\alpha}=(1,0,0),(0,0,1), (1/3,1/3,1/3)$, which corresponds to the parallel, vertical, isotropic polarization cases respectively.
The oscillatory behavior displayed in Fig.~(\ref{f}) is related to
the stationary modes of the fields as a result of the superposition of the propagating incident and reflected modes.
The oscillations of both the tangential  and normal components of the field  lead to the different oscillatory behaviors of the decay rates of both
the atoms polarizable in the directions parallel and vertical to the
planar boundary.

\section{ Effects of vacuum fluctuations on atomic frequency estimation}
As we have demonstrated, the presence of a reflecting boundary may  protect the estimation precision of the initial parameters from the influence of the environment in certain circumstances. We may also wonder how it affects other parameters' estimation such as the frequency estimation. Now, our unknown parameter becomes the atomic frequency $\omega_0$.
 By using Eq.~(\ref{FQ}) and Eq.~(\ref{Bloch}), the quantum Fisher information of $\omega_0$ in the unbounded case and the bounded case can be calculated, after dropping the terms which are of higher order in terms of the fine structure constant in the effective level spacing of the atom, to the first order as
\bea
F_{\omega_0}^{(0)}\approx\sin^2\theta\, e^{- \gamma_0\tau}\tau^2+16\bigg(\frac{\partial\gamma_0}{\partial\omega_0}\bigg)^2\bigg[1+
\frac{\sin^2\theta}{4}+\cos\theta\,(\cos\theta+2)\,e^{- \gamma_0\tau}\bigg]e^{- \gamma_0\tau}\tau^2\label{omega0}
\eea
and
\bea
F_{\omega_0}\approx &&\sin^2\theta\, e^{- \gamma_0\tau(1-f(\omega_0,z_0))}\tau^2+\nonumber\\
&&16\bigg\{\frac{\partial[\gamma_0(1-f(\omega_0,z_0))]}{\partial\omega_0}\bigg\}^2\times
\bigg[1+\frac{\sin^2\theta}{4}+\cos\theta\,(\cos\theta+2)\,e^{- \gamma_0\tau(1-f(\omega_0,z_0))}\bigg]\times\nonumber\\
&&e^{- \gamma_0\tau(1-f(\omega_0,z_0))}\tau^2\;.\label{omega}
\eea
So, the maximum quantum Fisher information is obtained when $\theta=\frac{\pi}{2}$, which is given, after dropping the higher order term,
by
\beq
F_{\omega_0}^{(0)}\approx e^{- \gamma_0\tau}\tau^2
\eeq
and
\beq
F_{\omega_0}\approx e^{- \gamma_0\tau(1-f(\omega_0,z_0))}\tau^2\;.
\eeq
Applying the equation $\frac{\partial F_{\omega_0}}{\partial\tau}=0$, we find that the maxima of $F_{\omega_0}^{(0)}$ and $F_{\omega_0}$, $\frac{4}{\gamma_0^2}e^{-2}$ and $\frac{4}{\gamma_0^2[1-f(\omega_0,z_0)]^2}e^{-2}$,  can be reached at $\tau^{(0)}=\frac{2}{\gamma_0}$ and $\tau=\frac{2}{\gamma_0(1-f(\omega_0,z_0))}$. When $z_0\rightarrow\infty$, $f_i(\omega_0,z_0)\rightarrow0$, the bounded vacuum case reduces to the unbounded case as expected. When $z_0\rightarrow0$, $f_x(\omega_0,z_0)=f_y(\omega_0,z_0)=-f_z(\omega_0,z_0)=1$, thus different polarization directions lead to different results. When the polarization direction is along the $z$ axis, i.e., $\alpha_x=\alpha_y=0$, the optimal measurement time and the maximal quantum Fisher information in the bounded vacuum case become $\tau=\frac{1}{\gamma_0}$ and $F_{\omega_0}=\frac{1}{\gamma_0^2}e^{-2}$. The optimal measurement time is half of that in the unbounded vacuum case and the maximal quantum Fisher information is a quarter of that in the unbounded case. When the polarization direction is in the $xy$ plane, i.e., $\alpha_z=0$, both the optimal measurement time and the maximal quantum Fisher information in the bounded case approach to infinity, and the quantum Fisher information with time $\tau$ approaches to $\tau^2$, which is consistent with the quantum Fisher information in unitary evolution $\bm{\omega}=(\cos(\phi+\omega_0\tau),\sin(\phi+\omega_0\tau),0)$. When polarization direction is isotropic, $\alpha_x=\alpha_y=\alpha_z=\frac{1}{3}$, the optimal measurement time and the maximal quantum Fisher information in the bounded vacuum case become $\tau=\frac{3}{\gamma_0}$ and $F_{\omega_0}=\frac{9}{\gamma_0^2}e^{-2}$, which are both larger than those in unbounded vacuum case. For an arbitrary polarization, in regions where  $f(\omega_0,z_0)>0$, both the optimal measurement time and the maximal quantum Fisher information are enhanced as compared to the unbounded vacuum case, while in regions where  $f(\omega_0,z_0)<0$, they are depressed.

Let us note that in estimation of a transition frequency, the total time $T$ of probing is also a resource. For a given $ T\gg{1\over \gamma_0}$,
we can improve the precision by accomplishing a sequence of measurements~\cite{Huelga}. We let $N=\frac{T}{\tau}$ denote the repeated times of the measurements. Then, according to Eq.~(\ref{Un}), the uncertainty of the atomic frequency satisfies
\beq
|\Delta\omega_0|\geq\frac{1}{\sqrt{N F_Q}}=\frac{1}{\sqrt{T F_Q/\tau}}=\frac{1}{\sqrt{T\tau e^{- \gamma_0\tau(1-f(\omega_0,z_0))}}}\;
\eeq
Applying $\frac{\partial |\Delta\omega_0|}{\partial\tau}=0$, we obtain the minimum uncertainty of $\omega_0$, i.e., $|\Delta\omega_0|_{min}=\sqrt{\frac{e\gamma_0(1-f(\omega_0,z_0))}{T}}$,
and the optimal sequence measurement time  $\tau=\frac{1}{\gamma_0(1-f(\omega_0,z_0))}$. As a result, the optimal repeated times of measurements in a given total time of probing to maximize the precision is  $T\gamma_0(1-f(\omega_0,z_0))$.
In regions where  $f(\omega_0,z_0)>0$, the precision is enhanced and the measurement times we need to obtain the maximum precision is less than that in the unbounded case. In regions where $f(\omega_0,z_0)<0$, the precision is degraded and the  measurement times become more than that in the unbounded case.
It could also be possible to further improve
this precision by the use of an entangled probe system following~\cite{Huelga},
the detailed analysis of which is left as future work.  Here, we give a very brief comment on the issue. Take $N$ maximally entangled atoms in vacuum for example 
and let $\rho_{tot}=\rho(0) \otimes |0\rangle\langle0|$  be the initial total density matrix of the system, where $\rho(0)$ is the initial reduced density matrix of the $N$ maximally entangled atoms. Then we  can use Eq.~(\ref{evo}) to study the evolution of the total state. After obtaining the state of the atoms in time,  we can in principle calculate the quantum Fisher information using Eq.~(\ref{FI}). Because of the indirect interaction between atoms caused by the field, it is not easy to find an analytical result of the reduced density matrix of the $N$ atoms in time $\tau$. But in the special case when all the coefficients of the dissipative part in the evolution equation of the reduced density matrix  vanish, all the atoms behave like closed systems and the Heisenberg precision limit is protected from deterioration. This condition can be approximately fulfilled if  atoms are transversely  polarizable  and very close to the boundary.

\section{conclusion}

In summary, we studied the dynamics of the quantum Fisher information for the atomic parameter estimation for a static polarizable two-level atom coupled in the multipolar scheme to a bath of fluctuating vacuum electromagnetic fields without and with the presence of a reflecting boundary.  When we estimate the parameters of initial atomic state, we find that, in the case without a boundary, the
 electromagnetic vacuum fluctuations always cause the quantum Fisher information of the initial parameters and thus the precision limit of parameter estimation  to decrease. However, with the presence of a boundary, the quantum Fisher information becomes position  and atomic polarization dependent, and thus the precision of the initial  parameter estimation may be decreased, enhanced or remain
 unchanged as compared to the case without a boundary depending on the position and polarization . When the atom is extremely close to
 the boundary and is transversely polarizable, the quantum Fisher information may even be shielded from the influence of the vacuum fluctuations and remains constant with time as if it were a closed system.
For the estimation of the atomic frequency, there exist a maximum quantum Fisher information and optimal measurement time, which can also both be enhanced or decreased as compared to the case without a boundary.

\begin{acknowledgments}
This work was supported by the National Natural Science Foundation of China under Grants  No. 11435006, and No. 11375092, and
the  Specialized Research Fund for the Doctoral Program of Higher Education under Grant No. 20124306110001
\end{acknowledgments}


\end{document}